\documentclass[journal=jpccck,manuscript=article]{achemso}
\usepackage{times}
\usepackage{epsfig}
\usepackage{graphicx}
\usepackage{psfrag}
\usepackage{ae}
\usepackage{amsmath,amssymb}
\usepackage[usenames]{color}
\usepackage{float}
\usepackage{xcolor}
\usepackage{placeins}
\usepackage{amsfonts}
\usepackage{bm}
\usepackage{setspace}
\usepackage{soul}
\usepackage{soulutf8}

\title{Dielectric behavior of Water in [bmim] [Tf$_2$N] room-temperature Ionic Liquid, molecular dynamic study.}

\author{ Ra\'ul Fuentes-Azcatl} 
\email{razcatl@correo.xoc.uam.mx}
\affiliation{Instituto de Qu\'{\i}mica, Universidade Federal Fluminense-Outeiro de S\~ao Jo\~ao Batista, s/n CEP:24020-141, Niter\'oi, RJ, Brazil}

\author{Gabriel J. C. Araujo\color{black}} 
\affiliation{Instituto de Qu\'{\i}mica, Universidade Federal Fluminense-Outeiro de S\~ao Jo\~ao Batista, s/n CEP:24020-141, Niter\'oi, RJ, Brazil}

\author{Tuanan C. Louren\c{c}o} 
\affiliation{Instituto de Qu\'{\i}mica, Universidade Federal Fluminense-Outeiro de S\~ao Jo\~ao Batista, s/n CEP:24020-141, Niter\'oi, RJ, Brazil}

\author{Cau\^{e} T. O. G. Costa} 
\affiliation{Instituto de Qu\'{\i}mica, Universidade Federal Fluminense-Outeiro de S\~ao Jo\~ao Batista, s/n CEP:24020-141, Niter\'oi, RJ, Brazil}

\author{Jos\'e Walkimar de M. Carneiro} 
\affiliation{Instituto de Qu\'{\i}mica, Universidade Federal Fluminense-Outeiro de S\~ao Jo\~ao Batista, s/n CEP:24020-141, Niter\'oi, RJ, Brazil}

\author{Luciano T. Costa} 
\email{ltcosta@id.uff.br}
\affiliation{Instituto de Qu\'{\i}mica, Universidade Federal Fluminense-Outeiro de S\~ao Jo\~ao Batista, s/n CEP:24020-141, Niter\'oi, RJ, Brazil}

\newpage 

\begin{document}
\date{}

\newpage

\maketitle
\begin{abstract}

In this work we present the dielectric behavior of water with a novel flexible model that improved all three sites water models Different concentrations of the ionic liquid 1-butyl-3-methylimidazolium [bmim] bis(trifluoromethanesulfonyl)imide [Tf$_2$N] with water was investigated. The study was performed by molecular dynamics simulations using three water models, being two non-polarizable 3-site SPC/E and SPC/e, and a novel flexible 3-site FAB/$\epsilon$ model. Systematic thermodynamics, dynamical and dielectric properties were investigated, such as density, diffusion coefficient, heat of vaporization $\Delta$Hvap, and surface tension at 300 K and 1 bar. We extrapolated the experimental molar fraction of the mixtures and a pattern change for all properties was observed, evidencing the phase separation previously reported by experimental data. The results also display the dielectric effect of the system on the calculated properties. 

\end{abstract}

\section{Introduction}

Ionic Liquids (ILs) have been investigated widely  in the last years, with applications being reported since the early 20th century. For instance, ILs were applied to cellulose dissolution in 1930s, when a pattent described the use of a molten pyridinium salt  for this purpose\cite{armand}.

Ionic liquids have great capacity to dissolve an extensive group of polar and non-polar compounds,making them a desirable solvent for many applications, such as catalysis, \cite{2,3,4} extraction, \cite{5,6}
self-assembly, \cite{7,8,9,10,11,12} electrochemical \cite{13,14,15} and biochemical
processes. \cite{16,17,18}

The search for new anions for organic polymer electrolytes raised the concept of a "plasticizing anion", i.e., an anion with delocalized charge and multiple conformations which differ in energy only slightly. One example of this is the anion bis (trifluoromethanelsulphonyl) imide [CF$_3$SO$_2$-N-SO$_2$CF$_3$]$^{\hspace{0.1cm}^{+}}$, also known as [Tf$_2$N]$^{\hspace{0.1cm}^{-}}$. It was shown that this anion, together with imidazolium cations such as [bmim][Tf$_2$N], forms an IL with a melting point around 288 K, ionic conductivity comparable to that of good organic electrolyte solutions \cite{armand} and no decomposition or significant vapour pressure up to 550-650K\cite{Earle}.

 It is well known that the water content can disrupt and strongly affect the ionic liquids properties ~\cite{ludwig2006,ludwig2009}. For instance  water has an important role in the dissolution of cellulose in ILs,  revealing the importance of hydrogen bonds making and disrupting environment. Recently, McDaniel and Verma studied the miscibility of ILs in water and showed the dependence of the dielectric medium based on ion type and cation/anion combination ~\cite{McDaniel2019}.  [bmim][Tf$_2$N] is not miscible with water, presenting a solubility around 1,000 p.p.m. in aqueous solutions\cite{armand}. Imidazolium based ILs and anions like [Tf$_2$N] form biphasic systems with water, being used in separation processes. ~\cite{coutinho2012} Due to that, the dielectric behavior of [bmim][Tf$_2$N] in water  becomes an interesting system to be explored. Even with some aspects being already investigated in the literature ~\cite{rollet}, others still remain.

A determining condition for the study of ionic systems in aqueous solution, within computational theoretical studies based on molecular dynamics technique, is the choice of the force field. Since the last century,  water force fields have been developed \cite{vega11,tip4pe,fab,spce}, being the TIP4P family one of the most used for improving their ability to tackle with the polarization effects, reproducing better the experimental values at various pressure and temperature conditions~\cite{vega11}. Since Vega et al \cite{vega11} proposed an equation which qualifies according to the experimental values and those reproduced by the model, it has been taken as a reference to qualify and develop new approaches for water force fields.\cite{tip4peM,fab}

As described by Vega et al\cite{vega11}, until 2011 the water models  failed with respect to the reproduction of dielectric properties that are determined by the dipole moment, inducing the development of various models since that year.\cite{tip4pe,spce,fab}, which has contributed to having better monovalent salt \cite{nacle,kbre}force fields. The FAB/$\epsilon$ model  reproduces more and better various experimental values~\cite{fab}. An advantage of the FAB/$\epsilon$ model over the other ones is related to its flexibility, helping to understand how the structure of water changes within a solution. In the present work we employ this model to study a water/IL mixture by the first time.  

In the literature there are several force fields for [bmim][Tf$_2$N], which have been established by the reproduction of some experimental property, such as density, X-ray structure factors, self-diffusion
coefficients from NMR and heats of vaporization from vapor pressure measurements\cite{kman}. The first force fields were obtained from ab initio \cite{lopes}  methods and compared with the reproduction of the liquid density. Currently there are more experimental data and therefore conditions to improve the IL force field.\cite{kman}

The remaining of the paper goes as follows. In section 2  we 
 introduce the models. Section 3 shows the simulation details.  The results are analysed in Section 4. Conclusions
are presented in section 5.
\newpage
\section{Computational Details and Models}
\subsection{Force Fields}
Three different Force Fields (FF) for water were used to investigate the effects on the ionic liquid properties, varying the water content. The SPC/E\cite{spcE}, a non-polarizable model which is widely applied, the SPC/$\epsilon$ \cite{spce} model, an improvement of SPC model with new Lennard-Jones and charges parameters, and the FAB/$\epsilon$ \cite{fab} model, a flexible 3-site model which can reproduce various thermodynamic properties at different conditions and improve the 3-sites water model.

 All these models have 3 point charges centered on each atom  and a Lennard-Jones site for the oxygen. 
Table \ref{table2} shows the force field parameters for the three models.

\begin{table}
\caption{Parameters of the three-site water models considered in this work.
}
\label{table2}
\begin{tabular}{|ccccccccc|}
\hline\hline
model	&	$k_{b}$	&	$r _{OH}$ 	&	$k_{a}$	&	$\Theta$ 	&	$\varepsilon_{OO}$	&	$\sigma_{OO}$	&	$q_{O}$	&	$q_{H}$	\\

	&	kJ/ $mol$ {\AA}$^{2}$ 	&	{\AA}	&	kJ/ $mol$ rad$^{2}$	&	deg	&	kJ/mol	
	&	{\AA}	&	e	&	e	\\

SPC/E \cite{spcE}	&	-	&	1.000	&	-	&	109.45	&	0.650155	&	3.1660	&	-0.8476	&	0.4238	\\
SPC/$\epsilon$	\cite{spce}&	-	&	1.000	&	-	&	109.45	&	0.705859	&	3.1785	&	-0.8900	&	0.4450	\\

FBA/$\epsilon $	\cite{fab}&	3000	&	1.027	&	383	&	114.70	&	0.792324	&	3.1776	&	-0.8450	&	0.4225	\\

\hline
\end{tabular}
\end{table}

In general, non-polarizable models have the Lennard-Jones and Coulomb potentials to describe the intermolecular interactions  as follows:
\begin{equation}
\label{ff}
u(r) = 4\epsilon_{\alpha \beta} 
\left[\left(\frac {\sigma_{\alpha \beta}}{r}\right)^{12}-
  \left (\frac{\sigma_{\alpha \beta}}{r}\right)^6\right] +
\frac{1}{4\pi\epsilon_0}\frac{q_{\alpha} q_{\beta}}{r}
\end{equation}

\noindent where $\alpha$ and $\beta$ refers to atomic sites, $r$ is the distance between the sites $\alpha$ and $\beta$, $q_\alpha$ and $q_\beta$ accounts to the electric charges for the sites $\alpha$ and $\beta$, respectively;  $\epsilon_0$ is the vacuum permittivity,  $\epsilon_{\alpha \beta}$ is the LJ energy scale and  $\sigma_{\alpha \beta}$ is the repulsive diameter for an $\alpha-\beta$ pair. The cross interactions between different atoms are obtained using the Lorentz-Berthelot mixing rules,

\begin{equation}
\label{lb}
\sigma_{\alpha\beta} = \left(\frac{\sigma_{\alpha\alpha} +
  \sigma_{\beta\beta} }{2}\right);\hspace{1.0cm} \epsilon_{\alpha\beta} =
\left(\epsilon_{\alpha\alpha} \epsilon_{\beta\beta}\right)^{1/2}
\end{equation}

The FAB/$\epsilon$ model includes two harmonic potential,that improve the calculation of experimental properties \cite{rfaCO2}, one at the OH bond and another at the angle formed by the three atoms of a water molecule, as shown in the Equations \ref{k}  and \ref{theta}:
 \begin{equation}
\label{k}
U_k(r)=\frac{k_r}{2}(r-r_0 )^2 
\end{equation}
 \begin{equation}
\label{theta}
U_{\theta}(\theta)=\frac{k_{\theta}}{2}(\theta-\theta_0)^2 ,
\end{equation}
\noindent where $r$ is the bond distance and $\theta$ is the bond 
angle. The subscript $0$ denotes their equilibrium
 values, while $k_r$ and $k_{\theta}$ are the corresponding spring constants. 

\begin{figure}
\includegraphics[scale=0.5,bb=14 14 704 253]{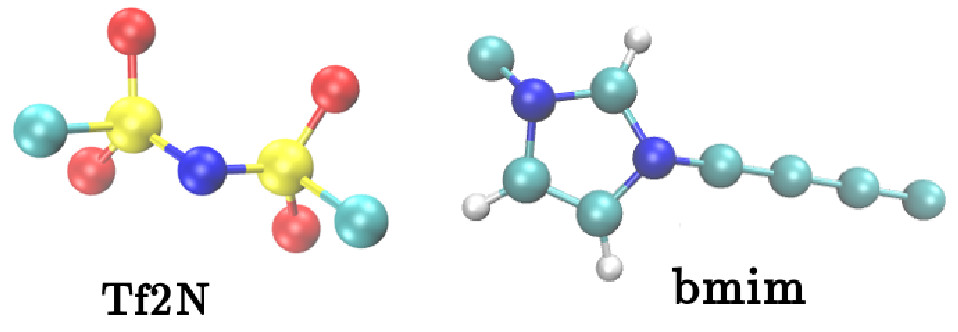}
\caption{The cyan spheres represent a Carbon atom, red spheres a Oxygen atom, blue spheres a Sodium atom, yellow spheres a Sulfur atom and white spheres Hidrogen (only the carbons at the aromatic ring have an explicit hidrogen), the contribution of the other Hydrogen atoms is contained in the respective carbons, as well as the contribution of the Fluorine atoms.}
\label{FigA}
\end{figure}

Figure \ref{FigA} exhibits the molecule model for the ions in the IL [bmim][Tf$_2$N], with the cation on the rigth and the anion on the left side. For the IL, the united atom force field proposed by K$\ddot{o}$ddermann et. al\cite{kman} was used. 

In this FF,  hydrogen atoms are not explicitly placed and their contribution is included in the atom to which it is bound. However,  carbon atoms in the aromatic ring have an explicit hydrogen, taken into account by placing an atom with specific characteristics according to the position where it is found. Also the energetic contribution of the fluorine atoms bonded to the carbons in Tf$_2$N anion are included in the corresponding carbon atom.

\section{Simulation Details}

Molecular Dynamics (MD) simulations on the liquid phase for binary mixtures were performed using the isothermal-isobaric (\textit{NpT}) ensemble \cite{Tuckerman} to obtain the density, the dielectric constant, the self-diffusion coefficient and the heat of vaporization. For this purpose the Gromacs package was used\cite{gromacs}. 
A total of 864 molecules were used in all simulations performed. The leapfrog equations of motion were solved with a time step of 1 fs. Periodic boundary conditions were used in all the directions and the bond distances were kept rigid with the LINCS procedure,\cite{Hess1} when necessary.
The cutoff distance applied was equal to 1.0 nm for both the real part of the Coulomb potential and the LJ interactions; In addition, analytical long-range corrections were applied for the second one. The PME method \cite{Essmann} was applied to evaluate the reciprocal contribution with a grid of 0.34 nm for the reciprocal vectors with a spline of order 4. The Nos\'e-Hoover thermostat and Parrinello-Rahman barostat were applied with the coupling times of 0.6 and 1.0 ps, respectively. The average properties were obtained after a production runs of 50 ns long.

The water static dielectric constant, $\epsilon$, is a collective property of an ensemble of water dipoles, which can be calculated from the equilibrium total dipole moment fluctuations, $(<\textbf{M}^2>-<\textbf{M}>^2)$.
The calculations of the dielectric constant was obtained by  (equation \ref{Ec4})~\cite{Neumann} of the total dipole moment {\bf M},

\begin{equation}
\label{Ec4}
\epsilon=1+\frac{4\pi}{3k_BTV} (<\textbf{M}^2>-<\textbf{M}>^2)
\end{equation}
\noindent where  $k_B$ is the Boltzmann constant, $T$ is the absolute temperature and $V$ is the volume.

The surface tension was obtained in a NVT simulation where a rectangular cell was used, with L$_x$=L$_y$=7.9 nm and L$_z$ = 3L$_x$ to avoid finite effects \cite{Orea}, containing 5324 molecules. Periodic boundary conditions were applied in all directions.

The average components of the pressure tensor were obtained
for 30 ns after an equilibration period of 5 ns. The densities of the
two phases were
extracted from the statistical averages of the liquid and vapor 
limits of the density profiles~\cite{alj95}. The corresponding surface 
tension $\gamma$ on  the planar interface was calculated 
from the mechanical definition of $\gamma$ ~\cite{alj95}.
\begin{equation}
\gamma=0.5L_z[<P_{zz}>-0.5(<P_{xx}>+<P_{yy}>)]
\end{equation}
where $L_z$ is the length of the simulation cell in the longest direction and $P_{\alpha\alpha}$ are the diagonal components of the pressure 
tensor. The factor 0.5 outside the squared brackets takes into account the two symmetrical interfaces in the system.

\section{Results}

\begin{figure}
\includegraphics[scale=0.5,bb=14 14 684 505]{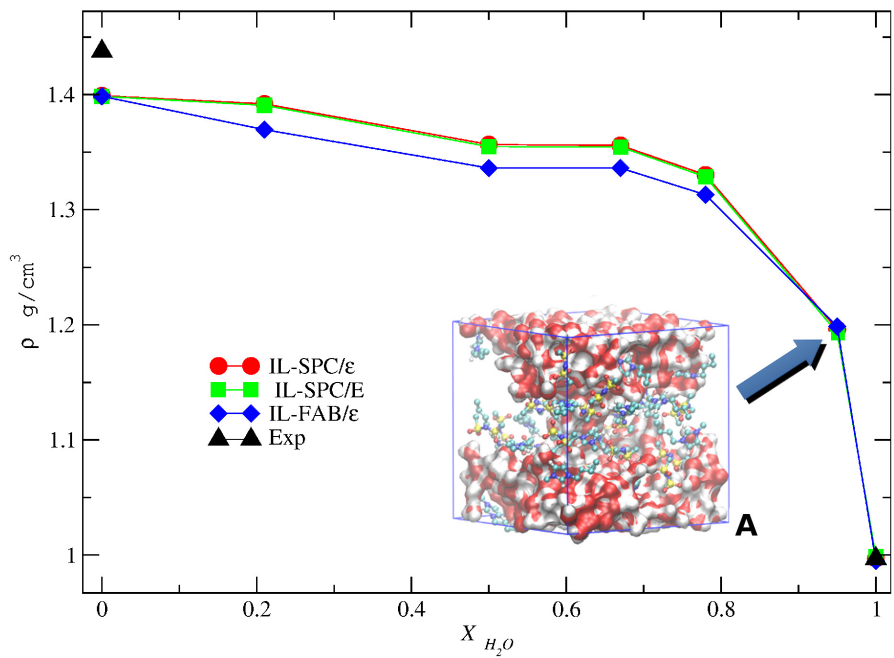}
\caption{Density of the water-IL solution as function of the molar fraction of water at 1 bar of pressure. Black triangles are the experimental data\cite{densIL}, red circles are the result using the SPC/$\epsilon$ model, green box are the result using the SPC/E model, and the blue dimonds are the result using FAB/$\epsilon$ model, all calculated in this work. }
\label{Fig1}
\end{figure}

 Figure \ref{Fig1} shows the density of the IL-water system at different mole fraction of water $X_{H2O}$. Although the calculated density of the pure IL is underestimated in 2.68\%, indicating that the force field for the IL can be improved, the behavior of the density using the three different force fields for water has a systematic decreasing. The inset (A) of the Figure 2 shows the equilibrated configuration at the fraction of $X_{H2O}$=0.95, where a non-homogeneous mixture is evidenced, suggested previously by Rollet et al\cite{rollet}. 

The density of the IL-water system (figure \ref{Fig1}) with respect to an increment of $X_{H2O}$ decreases by 40\% until reaching a composition of pure water; the density obtained with the FAB/$\epsilon$ model shows lower values than those obtained when using the non-polarizable models, within the region from $X_{H2O}$=0.2 to $X_{H2O}$=0.8. This is due to the contracting ability of the OH bond in the water molecule, as will be analyzed and discussed in the average structure of the water molecules in the system IL-water in this work.

\begin{figure}
\centerline{\psfig{figure=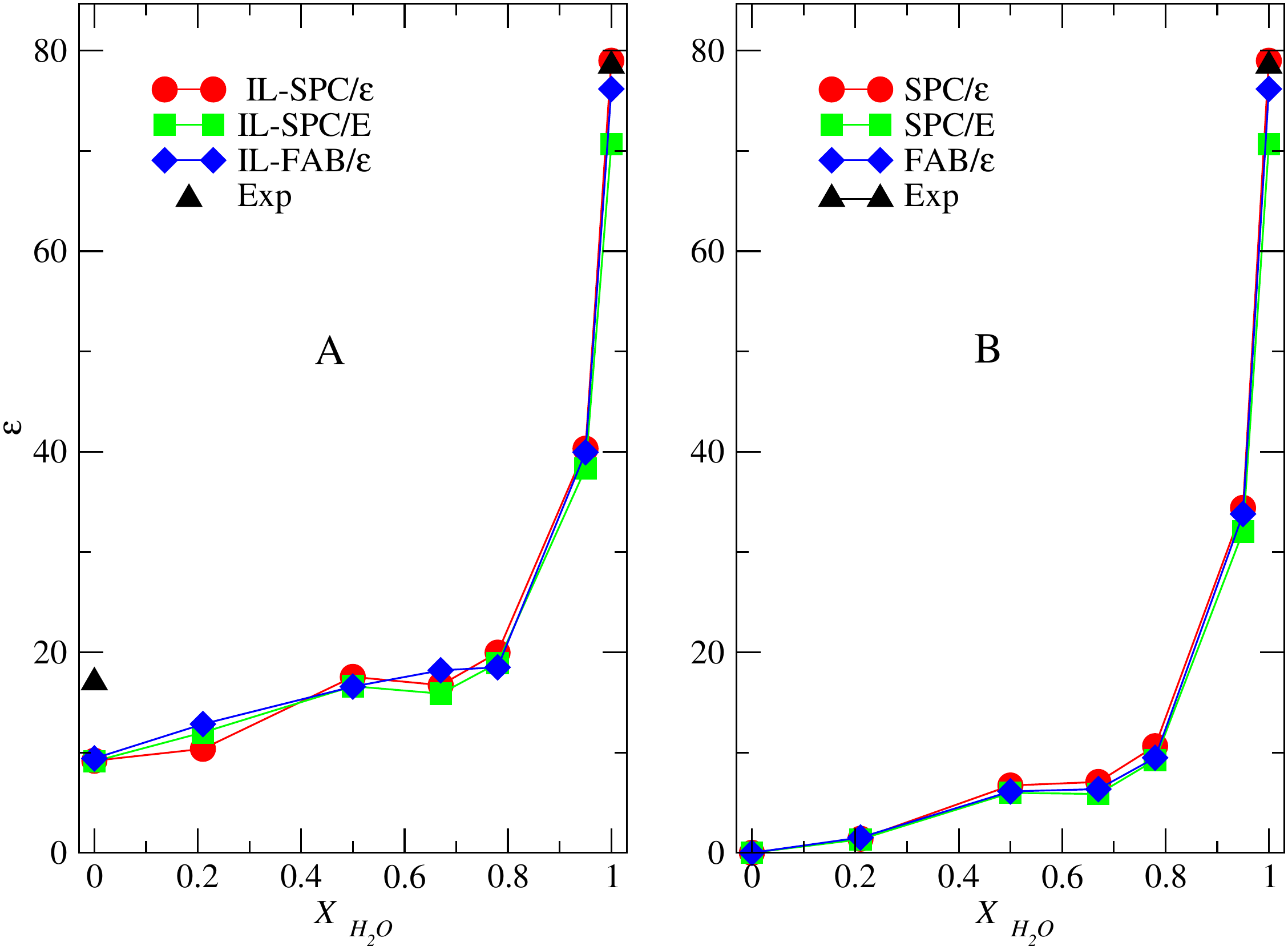,width=12.0cm,angle=0}}
\caption{\textbf{ A}: Dielectric constant of the water-IL solution as function of the molar fraction of water.\textbf{ B}: Dielectric constant of the water in the IL-water solution as function of the molar fraction of water. Both at 1 bar of pressure and 298 K of temperature.
Black triangles are the experimental data\cite{nist}, red circles are the result using the SPC/$\epsilon$ model, green box are the result using the SPC/E model and the blue dimonds are the result using FAB/$\epsilon$ model, all calculated in this work. }
\label{Fig2}
\end{figure}

Figure \ref{Fig2}A exhibits the results of the dielectric constant of the IL-water system for the three models studied. The difference between the water models is closer to the pure water, where the dielectric derived models present a better agreement to the experimental data.  Following the previous trend, the IL force field understimate the dielectric constant of the neat IL. The dielectric constant of the system can be reduced up to 60\%, when compared to pure water, even with a small amount of the IL. However, the system has a monotonic behavior for molar fractions of water lower than 80\%. Figure \ref{Fig2}A shows the dielectric constant of the IL in the system.

\begin{figure}
\centerline{\psfig{figure=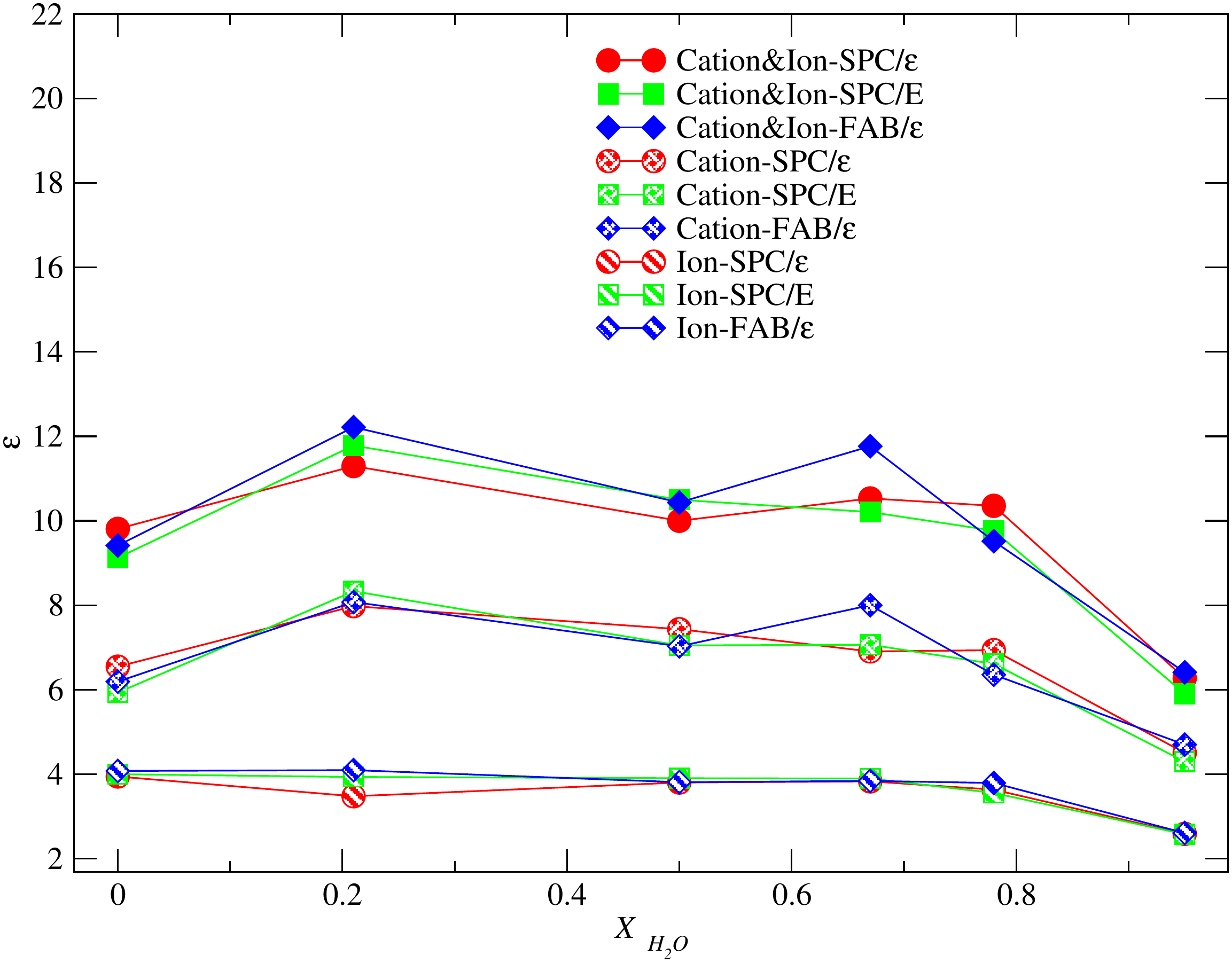,width=12.0cm,angle=0}}
\caption{Dielectric constant of the IL, cation and anion in the water-IL solution as function of the molar fraction of water at 1 bar of pressure.}
\label{Fig3}
\end{figure}

\newpage

Analizing the dielectric constant of water in the solution, Figure \ref{Fig2}B,   it can be seen  that the three water models show an identical description of this property. However, the FAB/$\epsilon$ model has the advantage of giving more information on the structural behavior of figures\ref{Fig4b}, \ref{Fig4c} and \ref{Fig4d}.

Figure \ref{Fig3} shows the behavior of the anion, cation and both in the IL-water system. The anion remains within a constant range, even when the amount of water in the system increases and the cation presents variations within a percentage of 20\% with respect to the absence of water. Thus the cation with this model has more reactivity than the anion.

The static dielectric constant, $\varepsilon$, of a polar liquid is related
to the thermal equilibrium fluctuations of the polarization
at zero field.\cite{fro} Polarization fluctuations are long-range and vary
with the shape of the dielectric body. $\varepsilon$, on the other hand, is an
intrinsic response coefficient independent of geometry. This led
Kirkwood to postulate that it should be possible to express $\varepsilon$ in
terms of a short-range orientational correlation function.\cite{kirk} 

\begin{figure}
\centerline{\psfig{figure=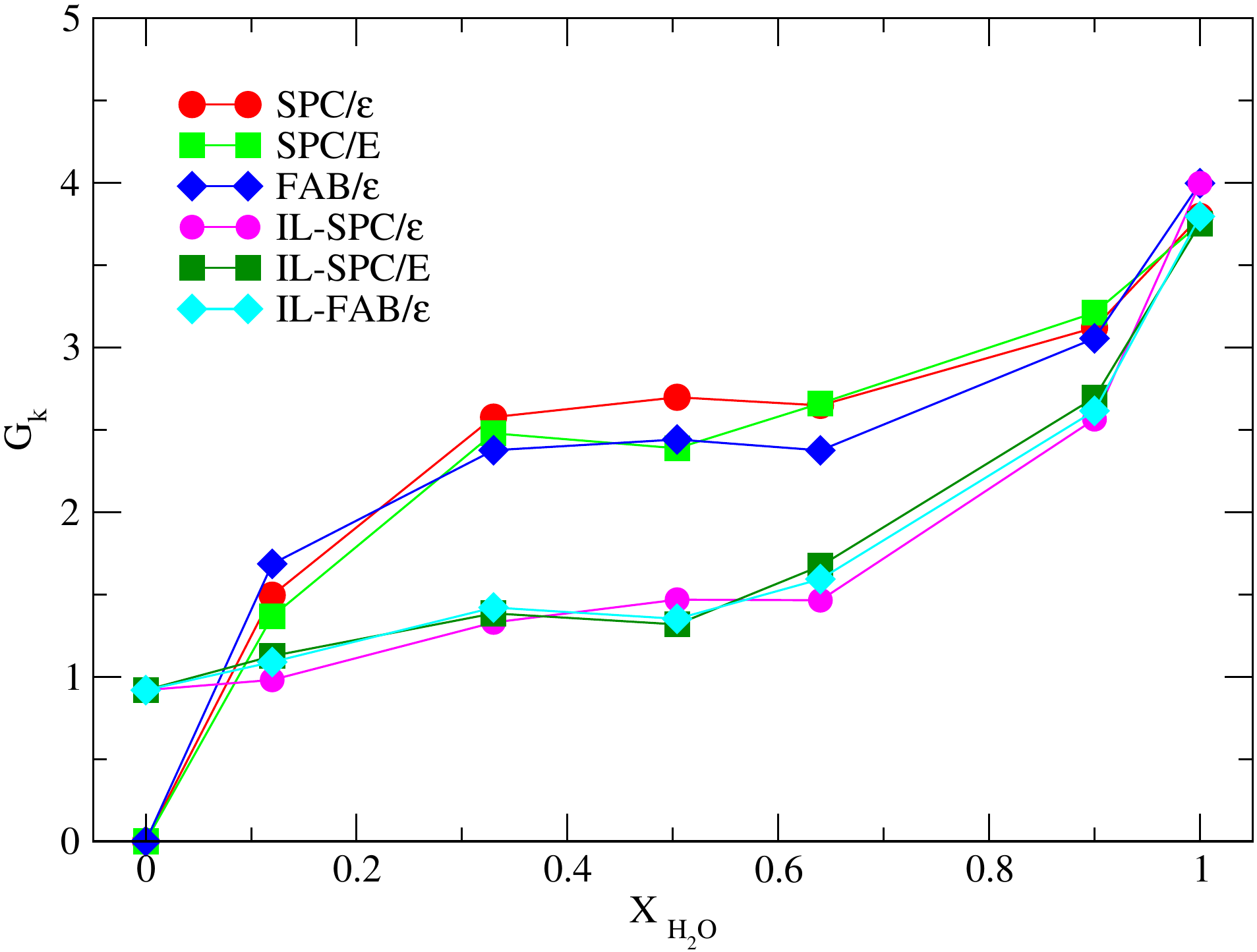,width=12.0cm,angle=0}}
\caption{Kirkwood factor G$_K$ of the water in the IL-water solution and for all solution as function of the molar fraction of water at 1 bar of pressure and 300K. Red circles are the result using the SPC/$\epsilon$ model, green box are the result using the SPC/E model and the blue dimonds are the result using FAB/$\epsilon$ model, all calculated in this work. }
\label{Fig4a}
\end{figure}

\begin{figure}
\centerline{\psfig{figure=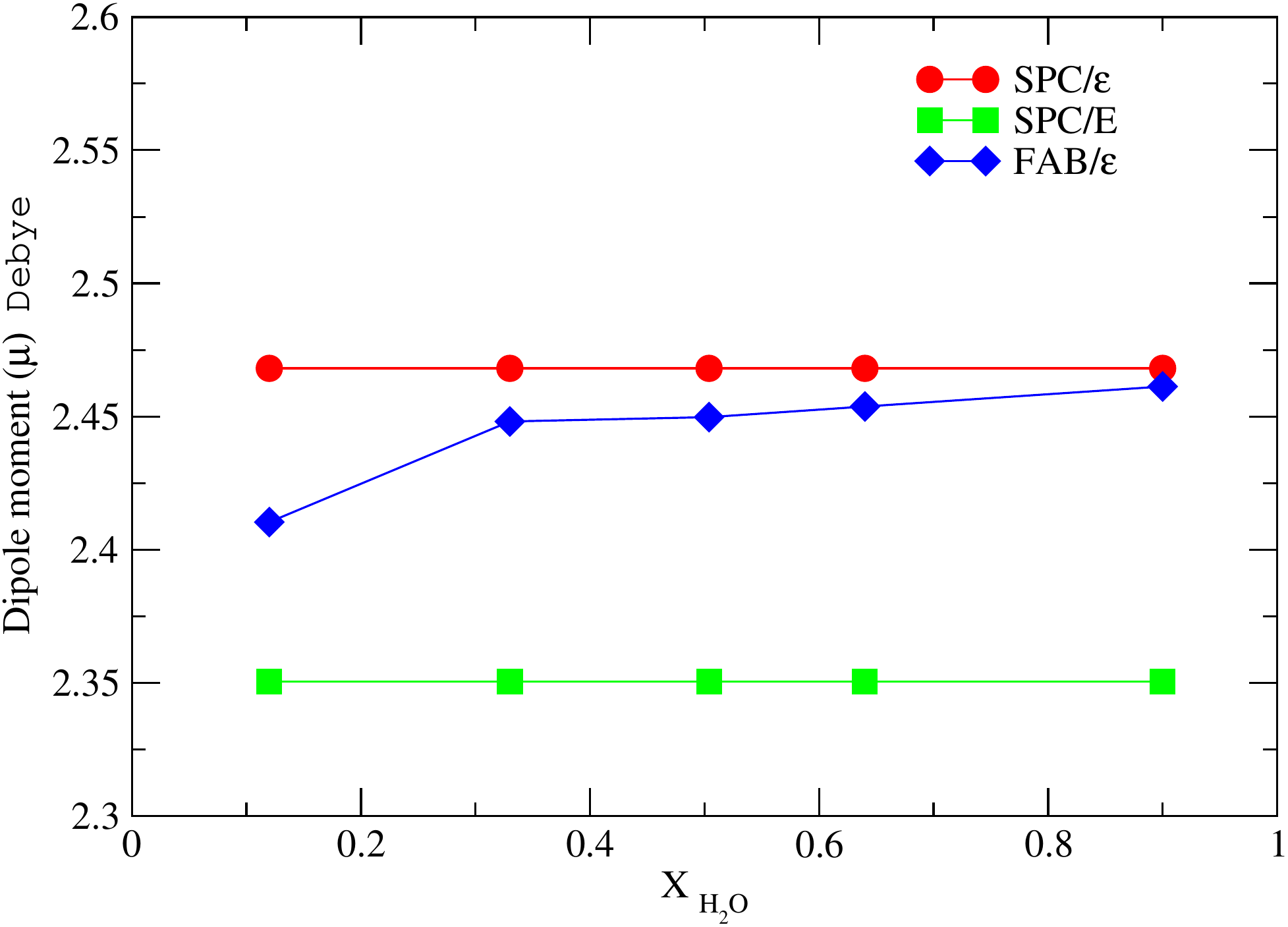,width=12.0cm,angle=0}}
\caption{Dipole Moment of the water in the IL-water solution as function of the molar fraction of water at 1 bar of pressure and 300K. Red circles are the result using the SPC/$\epsilon$ model, green box are the result using the SPC/E model and the blue dimonds are the result using FAB/$\epsilon$ model, all calculated in this work. }
\label{Fig4b}
\end{figure}

\begin{figure}
\centerline{\psfig{figure=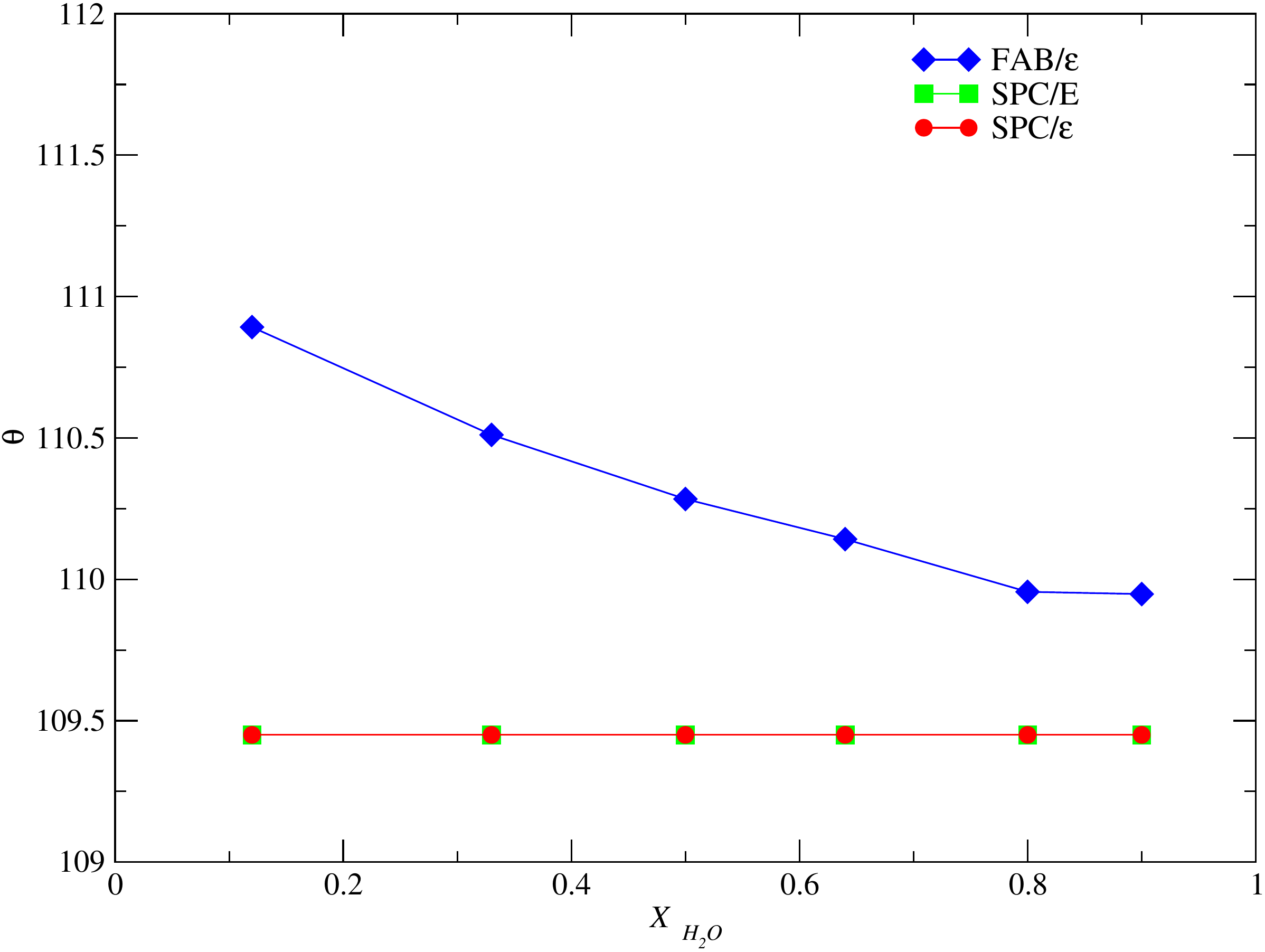,width=12.0cm,angle=0}}
\caption{Average angle of the water described by the three force fields as function of the molar fraction of water at 1 bar of pressure and 300K. Red circles are the result using the SPC/$\epsilon$ model, green box are the result using the SPC/E model and the blue dimonds are the result using FAB/$\epsilon$ model, all calculated in this work.}
\label{Fig4c}
\end{figure}

\begin{figure}
\centerline{\psfig{figure=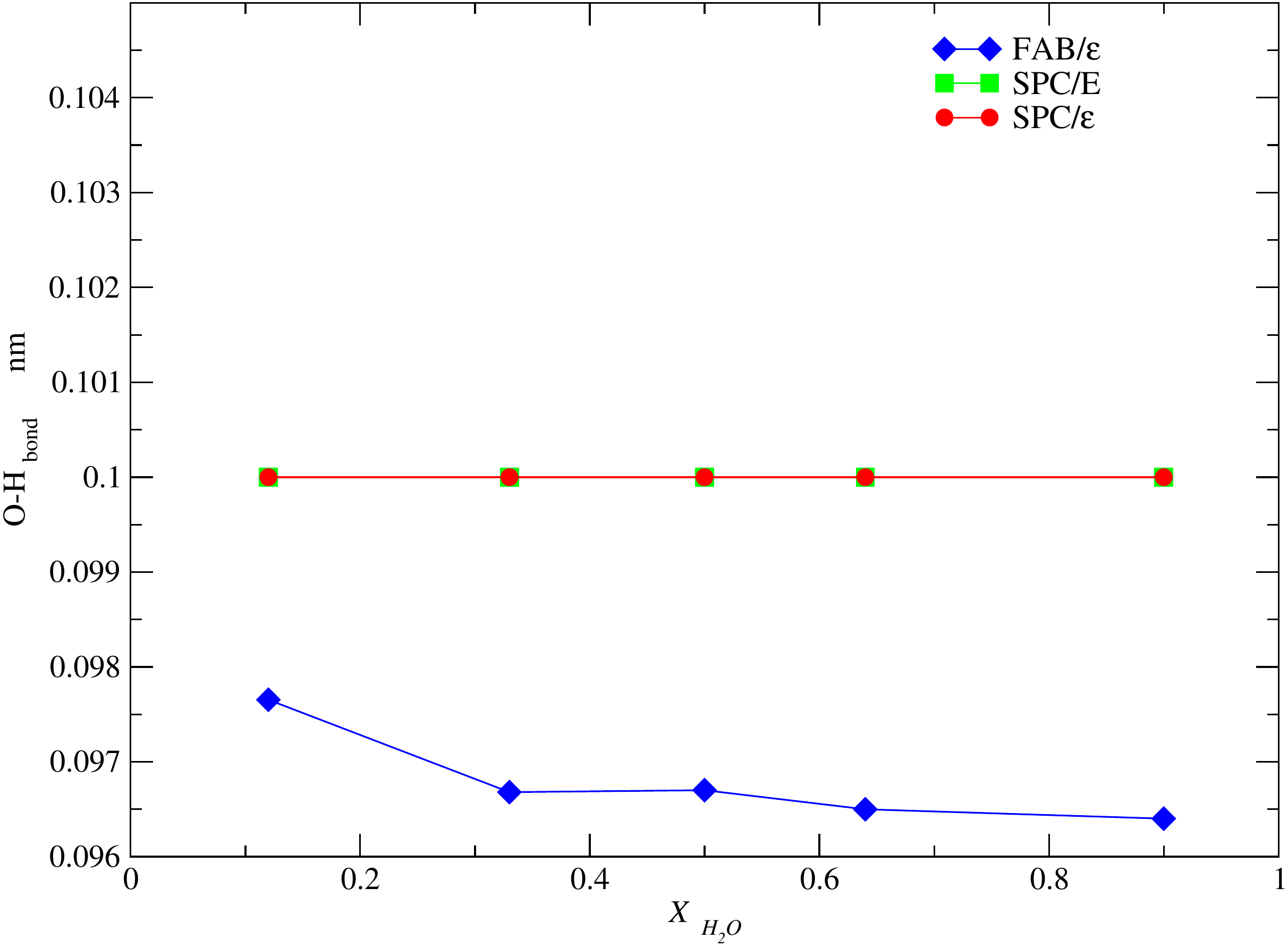,width=12.0cm,angle=0}}
\caption{Average O-H$_{bond}$ of the water described by the three force fields as function of the molar fraction of water at 1 bar of pressure and 300K. Red circles are the result using the SPC/$\epsilon$ model, green box are the result using the SPC/E model and the blue dimonds are the result using FAB/$\epsilon$ model, all calculated in this work.}
\label{Fig4d}
\end{figure}

The values of the polarization  of simulation is presented in the figure \ref{Fig4a}. We introduce the polarisation factor G$_K$ (equation \ref{Ec5})\cite{Glattli}, in order to check if the force field changes its polarisation; the G$_K$  factor measures the equilibrium fluctuations of the
collective dipole moment of the system and  is related to the orientational correlation function.. The calculus of the polarization is consistent for the three water models, including the polarizable FAB/$\varepsilon$ model, which implies that the system is consistently reproduced by these water models, and that water also governs the polarization of the system. By having a small amount of water in the system, it becomes more polarized and this causes its dielectric properties to increase, even when the system is not miscible.

The result is consistent with the three water models, Figure \ref{Fig4a}, which indicates that a water model that does not reproduce the dielectric constant well, will not reproduce this behaivor, as shown in the work of Schr\"oder et al \cite{ Schroder}
\begin{equation}
\label{Ec5}
G_K= <{\bf M}^2> / N \mu^2
\end{equation}

N is the total number of molecules, M is the total sum of dipoles $ {\mu} _i$ in the system (including the dipole of initial molecule ). Local orientational correlations are averaged out by thermal motion
after the first few coordination shells.


Even though the result of the dielectric constant and the polarization of water in the system with IL is consistent, the use of the new force field generates structural information on the behavior of water through the dipole moment, as seen in the figure \ref{Fig4b}. The average dipole moment of water, when increase its quantity with respect to IL, show water changes structurally with an angle less than isolated water, when it is closer to the IL. And by not being in contact with IL, it again deploys to a state of bulk.
\newpage
The dipole moment is a function of the location of the charges in the molecules with respect to time. In the case of rigid molecules we have a restorative force that turns the molecule fixed; in the case of the FAB/$\epsilon$ model, we have the possibility of having a displacement with a certain freedom caused by the harmonic potential, both in the angle and in the OH bond, which makes the structure free. From there we can obtain a broad understanding of how the water molecule behaves through various concentrations within this IL study, figure \ref{Fig4b}.


The dipole moment increases with respect to the amount of water in the system, this is due to the fact that the angle decreases in this process, see figure \ref{Fig4c}, approaching the hydrogens in the molecule; however the bond distance between oxygen and hydrogen O-H $_{bond}$ also decreases, see figure \ref{Fig4d}, which makes the interaction in the molecule stronger due to the contraction it undergoes. Then, when it reaches the bulk of pure water the molecule is more compact and when it is in a small quantity with ionic liquid it relaxes expanding its angle and O-H bond $_{bond}$
\newpage
\newpage

\begin{figure}
\centerline{\psfig{figure=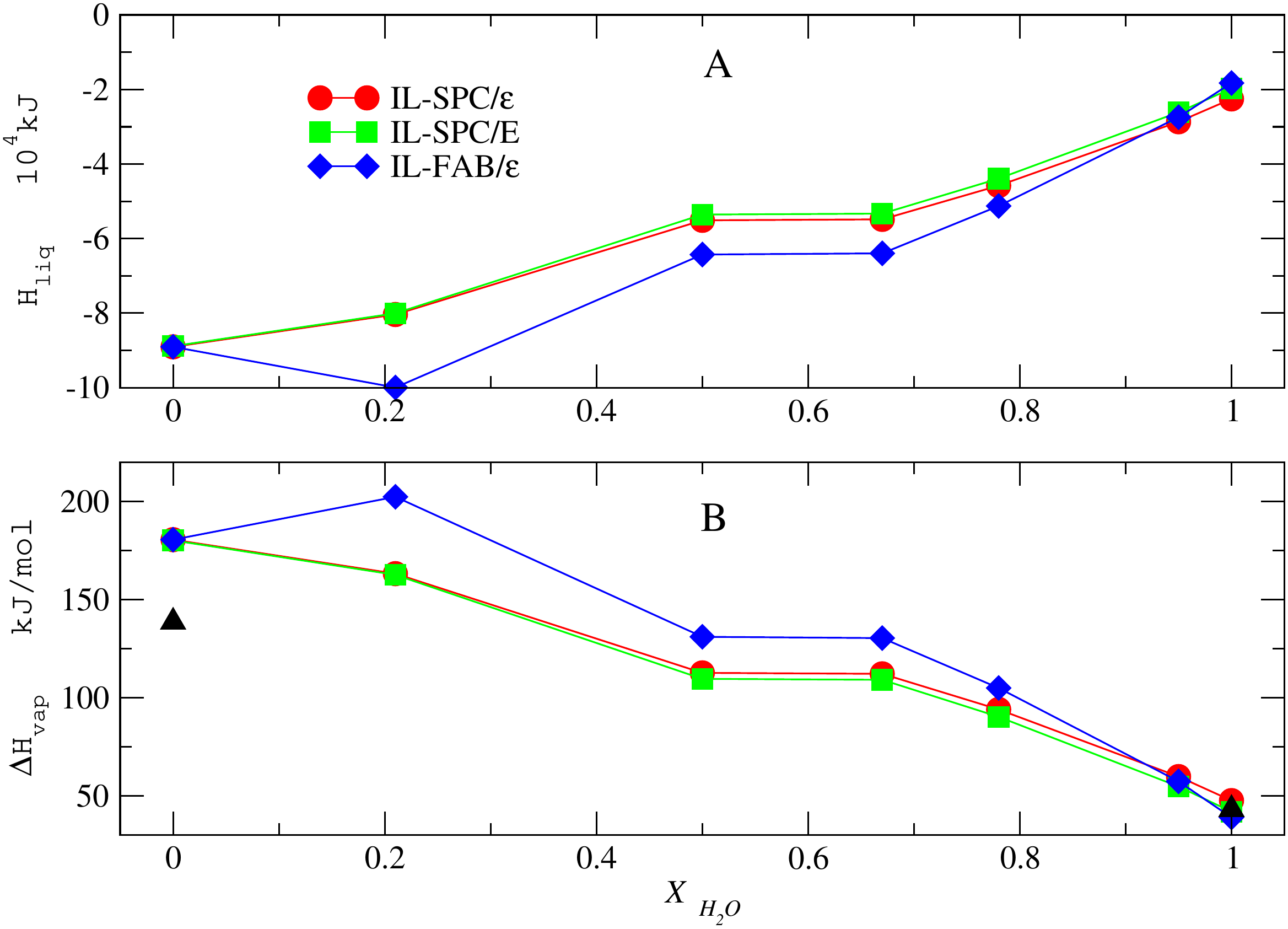,width=12.0cm,angle=0}}
\caption{\textbf{A}. Enthalpy of the liquid phase the IL-water solution as function of the molar fraction of water.
\textbf{B}. $\triangle$Hvap of the system with respect to the fraction of water in the solution IL-water. Both at 1 bar of pressure and 300K of temperature.  Black triangles are the experimental data\cite{Zaitsau}, red circles are the result using the SPC/$\epsilon$ model, green box are the result using the SPC/E model and the blue dimonds are the result using FAB/$\epsilon$ model, all calculated in this work. }
\label{Fig5}
\end{figure}
\newpage
The analysis of the enthalpy of the liquid phase (Figure \ref{Fig5}A), as we increase the amount of water in the solution, shows a decrease that stabilizes within a composition between 0.5 to 0.64 of the fraction of water. Outside that range there is a linear increase in the enthalpy.


The $\triangle$Hvap of the system(Figure \ref{Fig5}B) has a linear behavior outside the range of 0.5 to 6.4 of the fraction of water. Since the force field of IL underestimates the value of $\triangle$Hvap of pure IL, the value calculated with the FAB water model might seem wrong, but having a calculation of pure water closer to experimental allows us to infer that the work must focus on a better force field of the IL , in order to have a better description of this property.

\begin{figure}
\centerline{\psfig{figure=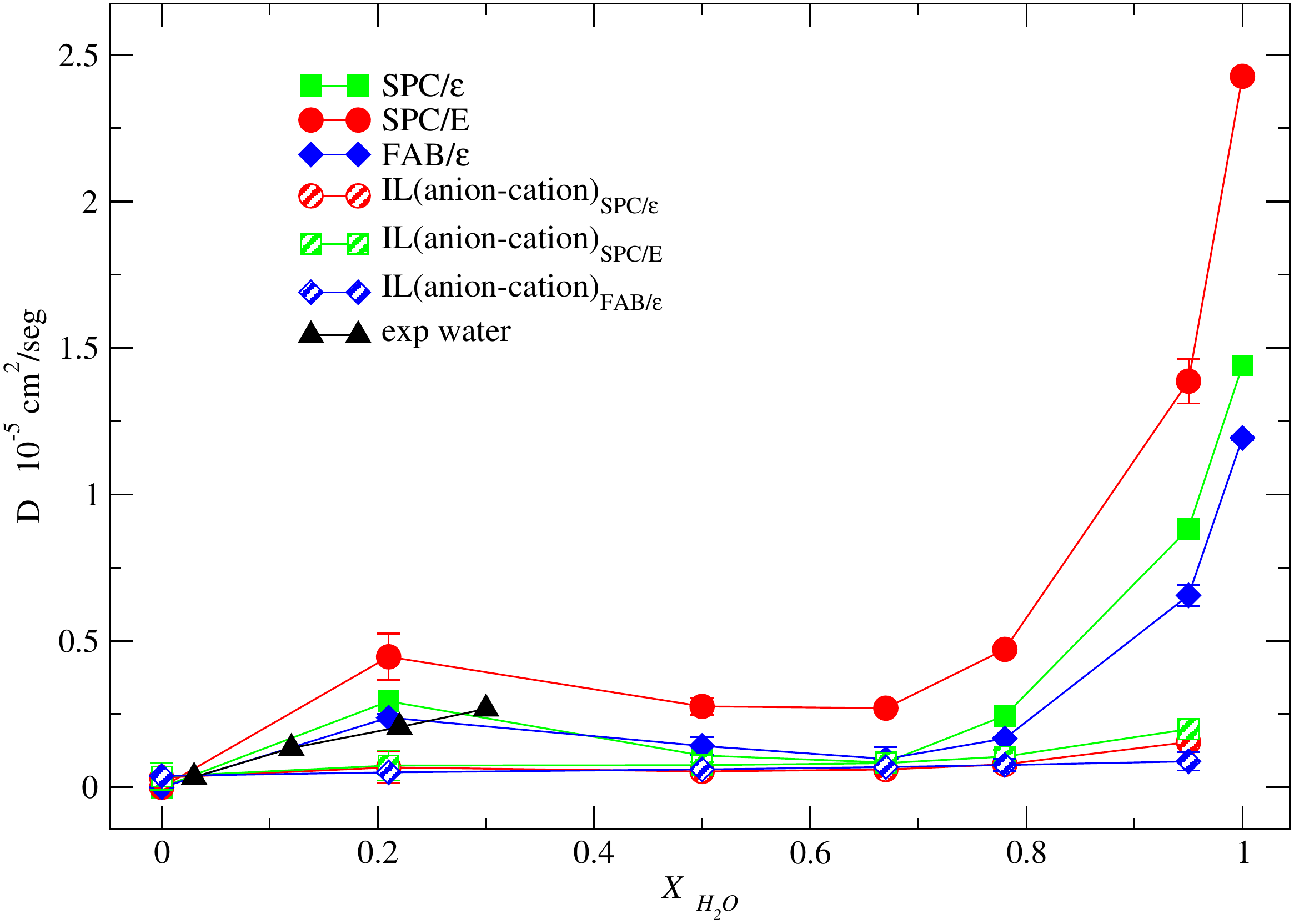,width=12.0cm,angle=0}}
\caption{ The self-diffusion coefficients of water and the IL with respect to the fraction of water in the solution IL-water. Black triangles are the experimental data\cite{rollet}, red circles are the result using the SPC/$\epsilon$ model, green box are the result using the SPC/E model and the blue dimonds are the result using FAB/$\epsilon$ model, all calculated in this work.  }
\label{Fig7}
\end{figure}
The self-diffusion coefficients of water is show in Figure \ref{Fig7}. When the water mol fraction is 0.2 in the solution, we have a description close to the experimental value of the diffusion, which describes the three water models, where the combination of IL-FAB/$\epsilon$ describes this behavior  closer to the experiment. Then there is a decrease indicated by the three systems until the system reach the value that every water model reproduce at 1 bar and 298K. Taking into account that IL is somewhat hygroscopic and not very miscible in water, the value it contains is 1.3\cite{rollet} percent by mass, at the pressure and temperature conditions used.

\newpage
\begin{figure}
\centerline{\psfig{figure=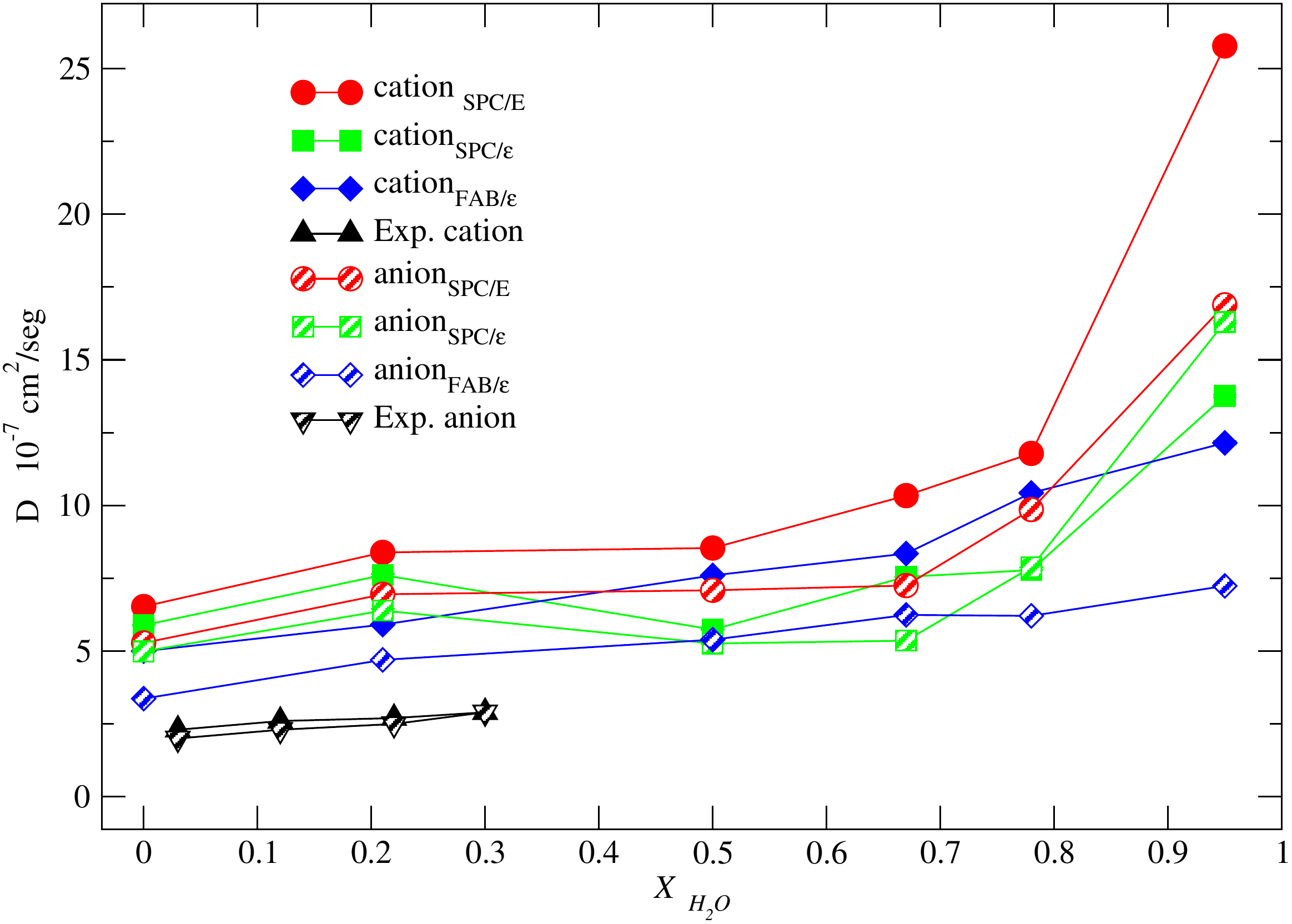,width=12.0cm,angle=0}}
\caption{ The self-diffusion coefficients of anion and the cation with respect to the fraction of water in the solution IL-water. Black triangles are the experimental data\cite{rollet}, red circles are the result using the SPC/$\epsilon$ model, green box are the result using the SPC/E model and the blue dimonds are the result using FAB/$\epsilon$ model, all calculated in this work. }
\label{Fig8}
\end{figure}
The self-diffusion coefficients of cation and anion is described by the Figure \ref{Fig8}. Although the diffusion is almost equal for both according to the reported experimental value \cite{rollet}, the calculated values that were obteined with the FBA/$\epsilon$ model are closer to the experimental values.
Self-diffusion coefficient result is interesting, since IL is expected to be a little hygroscopic, so that water increases the dissociation rate of the ion pair and consequently causes a higher diffusion increase mainly from the anion, due to its size. When the amount of water is high, the diffusion of both is greater by a factor 1.3 than when there is less
amount of water. These calculations show that increasing the water in the system does not produce a significant
increased dissociation of the anion-cation pair, that is, it appears
that water interacts rather weakly with the ion pair. Heinz et al \cite {heinz} discuss how the water content depends on the interaction with the anion. The hydrophobicity of anions [NTf$_2$] prevents the water from dissociating ion pairs and incorporates more between the IL.

\begin{figure}
\centerline{\psfig{figure=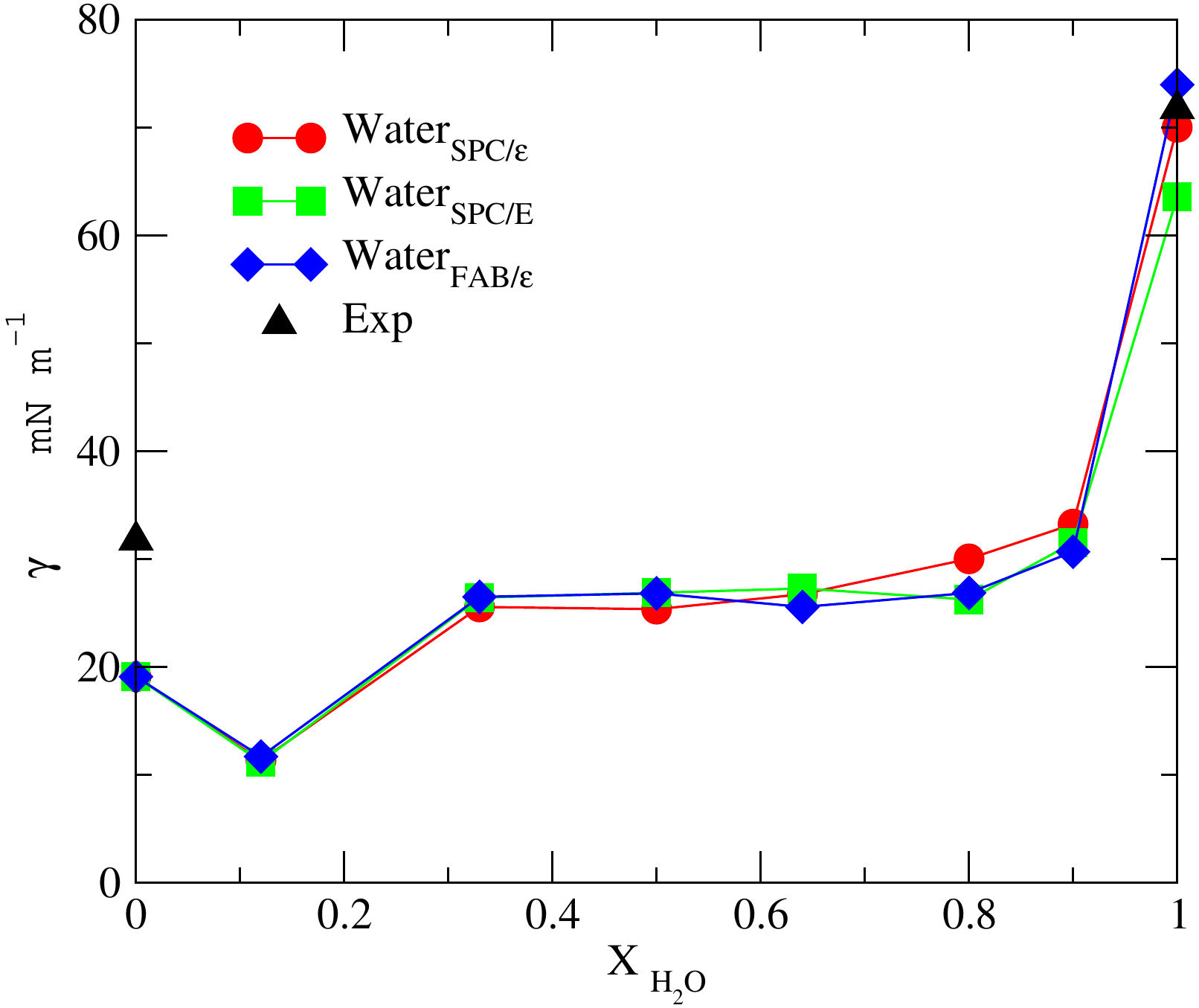,width=12.0cm,angle=0}}
\caption{ The  surface tension of the system with respect to the fraction of water in the solution IL-water. Black triangles are the experimental data\cite{rollet}, red circles are the result using the SPC/$\epsilon$ model, green box are the result using the SPC/E model and the blue dimonds are the result using FAB/$\epsilon$ model, all calculated in this work. }
\label{Fig9}
\end{figure}

The surface tension of the system when the water is increased is show in the Figure \ref{Fig9}. Although the SPC/E model diverges from the experimental value by almost 12\% at 300K and 1 bar, there is a similar behaivor since we started adding water to pure IL, until the increase reaches a mole fraction of 0.9 water. With a small amount of water, (  \begin{large}$\chi$\end{large}$_{H_{2}0}$= 0.12), the system decreases in value by almost 50\%, which may be attributable to the fact that the IL force field does not describe this property well.
From  \begin{large}$\chi$\end{large}$_{H_{2}0}$ = 0.32 to  \begin{large}$\chi$\end{large}$_{H_{2}0}$ = 0.8 we notice a similar behavior of the system, with small additions of water the system increases its value widely.

\newpage
\section{Conclusions}

The novel flexible water FAB gives a novel information on how water is structured in these systems and how its dipole moment changes, this helps to understand water with IL that does not solubilize, giving a guideline to improve IL. The three water models used in this work together with IL reproduce the experimental properties of the system from a small amount of water to pure water. The fact that the FAB/$\epsilon$ model can change its structure within the simulation, generating a type of polarization, allows us to have a  descriptive aspect of how the water molecule behaves through increasing its presence in the system, giving an  important description to what other force fields can not do.
This work shows the behavior and evolution of the three species [BMIM] [Tf$_2$N] + water at different concentrations, theoretically describing the behavior where there is still no experimental data but hypotheses have already been made regarding the system in these areas. 
An important challenge is to improve the IL force field in order to have a model closer to the experimental data and to have more certainty in the intermediate calculations of the properties calculated here.

The dielectric behavior of water within IL, as we increase its content, is reflected in the dipole moment, which indicates how the charges are located with respect to the atoms that form the molecule, and this is generated by the average structure defined by the O-H$_{bond}$ and angle, so the polarization of the water increases as its angle and O-H$_{bond}$ decrease.

\section {Acknowledgements}

Luciano T. Costa thanks the Brazilian Federal Agency for Support and Evaluation of Graduate Education (CAPES/Print/UFF grant number 88881.310460/2018-01) and also for the CNPq fellowship grant. RFA thank CAPES posdoc scholarships.

\end{document}